\documentclass[journal=apchd5,manuscript=article]{achemso}

\usepackage{chemformula} 
\usepackage[T1]{fontenc} 
\usepackage{url}
\usepackage{graphicx}
\usepackage{caption}

\usepackage[numbers,sort&compress]{natbib}

\usepackage{doi}
\usepackage{amsmath}
\usepackage{svg}
\usepackage{graphicx}
\usepackage[utf8]{inputenc}        
\usepackage[LGRgreek]{mathastext}
\usepackage{ragged2e}

\setkeys{acs}{doi=true}
 

\author{Sioneh Eyvazi}

\author{Evgeny A.~Mamonov}
\author{Rebecca Heilmann}
\author{Javier Cuerda}
\author{P\"{a}ivi T\"{o}rm\"{a}}
\email{paivi.torma@aalto.fi}
\affiliation{Department of Applied Physics, Aalto University School of Science, P.O. Box 15100, Aalto, FI-00076, Finland}

\title[Flat-Band Lasing in Silicon Waveguide-Integrated Metasurfaces]
  {Flat-Band Lasing in Silicon Waveguide-Integrated Metasurfaces}

\abbreviations{}
\keywords{flat\,band, lasing, silicon, metasurface, BIC, polarization vortex.}

\begin{document}

\begin{abstract}
Photonic flat\,bands are crucial for enabling strong localization of light and enhancing light-matter interactions, as well as tailoring the angular distribution of emission from photonic structures. These unique properties open pathways for developing robust photonic devices, efficient nonlinear optical processes, and novel platforms for exploring topological and quantum phenomena.
So far, experimental realizations of lasing in photonic flat\,bands have been limited to structures that emulate geometrically frustrated lattices in the tight-binding, i.e.~short-range coupling, regime. Here we consider a periodic metasurface with long-range couplings combined with guided modes and report experimental observation of lasing in photonic nearly flat modes. By carefully tuning the thickness of the guiding layer and periodicities, the observed flat lasing spectrum extends up to approximately $k_{y}= 2 \,  \mu m^{-1}$ in reciprocal space. Simulations show that the observed modes exhibit localization in both the waveguiding and active layers. In addition, we observe accidental bound states in the continuum (BICs) at the lasing frequencies, manifesting through polarization vortices with a topological charge |q|= 1.
 
\end{abstract}

\section*{Introduction}

Imposing a periodic potential for light has led to a vast amount of seminal discoveries since the advent of the distributed feedback laser \cite{kogelnik1971stimulated} and photonic crystals \cite{meade2008photonic}, boosted by modern design and fabrication techniques~\cite{ma2021deep}. In this context, photonic flat\,bands offer exciting and unique opportunities.  

A flat\,band refers to a mode where the optical frequency remains constant over all wave vectors (momenta) in a relevant range (e.g., the first Brillouin zone in the case of periodic systems), resulting in non-dispersive behavior. The flat\,band's high density of states (DOS) may enhance nonlinear effects, increase spontaneous emission, and lead to low threshold lasing. Furthermore, the energy (frequency) degeneracy of the wave vectors means that the system can be used for creating spacial distributions of light of essentially any type: both localized and extended. Finally, in two-dimensional structures where the in-plane momentum directly maps to the angle of light emitted from the structure, a flat\,band leads to angularly uniform emission or absorption, which can be interesting for some applications.  

In many photonic lattices light scatters from one unit cell and radiates over many others, realizing effectively long-range couplings: the bands emerge from the interference of the scattered light. Such long-range coupled photonic lattices may exhibit (nearly) flat\,bands of extremely low group velocity, demonstrating slow light properties for light traveling in the structure~\cite{settle2007flatband,altug2005experimental}.
This mechanism can lead to increased interaction between light and matter and enhanced control over light propagation due to the high DOS. The experimental observation of long-range coupled photonic flat modes has been extensively investigated in different structures such as photonic crystals and metasurfaces \cite{amedalor2023high,choi2024nonlocal,munley2023visible,nguyen2018symmetry,choi2024observation}.
In contrast, in condensed-matter physics, electronic flat\,bands often emerge in frustrated geometries through quantum tunneling between adjacent atoms, modeled by short-range coupled (tight-binding) Hamiltonians. This mechanism relies on the nearly localized nature of electrons in the crystal lattice such as twisted graphene, the kagome lattice, and the Lieb lattice~\cite{leung2020interaction,luan2023reconfigurable,harder2021kagome,leykam2018artificial,drost2017topological,andrei2020graphene,chen2023visualizing,slot2017experimental,tang2020photonic}. 
The rich physics of such systems has inspired photonic and exciton-polariton realizations of tight-binding Hamiltonians in frustrated lattices~\cite{mukherjee2015observation,baboux2016bosonic,scafirimuto2021tunable,masumoto2012exciton}. Flat\,band lasing and condensation have been observed in such flat\,bands that emulate tight-binding Hamiltonians~\cite{baboux2016bosonic,klembt2017polariton,whittaker2018exciton,harder2020exciton,harder2021kagome}. Our goal is to explore the possibilities of lasing in photonic structures that combine scattering from a periodic potential, and the resulting long-range couplings, with the opportunities of localization in and near the waveguiding layer offered by guided modes.

In this article, we experimentally observe lasing in the photonic flat\,bands hosted by an all-dielectric metasurface. Dielectric metasurfaces are promising for strong light-matter interaction due to their low optical loss and high refractive index. They have diverse applications in sensing, beam shaping, polarization detection, and imaging~\cite{tseng2020dielectric,chung2023dielectric,hu2020all,arbabi2015dielectric,li2020dielectric,wang2024miniaturized}. Notably, dielectric metasurfaces have been the subject of numerous studies focusing on lasing in different modes, particularly the bound state in the continuum (BIC)~\cite{azzam2020room,azzam2021single,ha2018directional,wu2020room}. However, 
lasing in a flat\,band has not been reported; as mentioned above, it has been observed only in tight-binding exciton-polariton lattices~\cite{harder2020exciton,harder2021kagome} (see also a numerical study~\cite{longhi2019photonic}).

We achieve lasing in nearly flat\,bands by employing a high refractive index amorphous silicon ($\alpha$-Si) waveguide-integrated metasurface. The $\alpha$-Si waveguide-integrated metasurface consists of a thin layer merged with a rectangular array of nanoparticles of the same material. This configuration hosts extended guided photonic modes, including an off-$\Gamma$ point accidental BIC, and nearly-flat modes, which we will refer to as flat\,bands for brevity. We combine the all-dielectric metasurface with an external gain medium, IR-140 dye solution. Through optical pumping, we experimentally observe lasing emission in the flat\,band. Remarkably, our eigenvalue simulation shows that the polarization of the flat mode is predominantly out-of-plane, making it advantageous for lasing since coupling with the emitter (gain medium) is well-defined and effective. In addition, we found a winding of the polarization field in the experiments, indicating a |q|= 1 topological charge at the accidental BIC points; this contributes to the active research on BICs in various photonic structures~\cite{kang2023applications,mohamed2022controlling,doeleman2018experimental,zhen2014topological,jin2019topologically,koshelev2019meta,kang2022merging,arjas2024high,ardizzone2022polariton,murai2020bound}. 

This article is organized so that we first describe the general structure of the samples, namely a thin waveguide layer overlaid with a nanoparticle array, both from the same dielectric material. We then discuss an example experimental structure showing two flat\,bands. We explain these bands by a simple analytical theory of the combined guided and lattice modes, and by numerical simulation of the electromagnetic field profiles in and near the nanoparticles and the waveguide layer. We also show how the flat\,bands can be modified by the lattice periodicity. The gain medium is then added, and we demonstrate the flat\,band lasing and analyze accidental BICs visible in the band structure and lasing emission. Finally, we present lasing in a structure optimized to produce a flat mode that extends over as large a momentum interval as we were able to realize with these structures. A summary and discussion of the results are presented in the conclusions.

\section*{Results and discussion}

\begin{figure}[hpt]
    \centering
    \includegraphics[width=15cm]{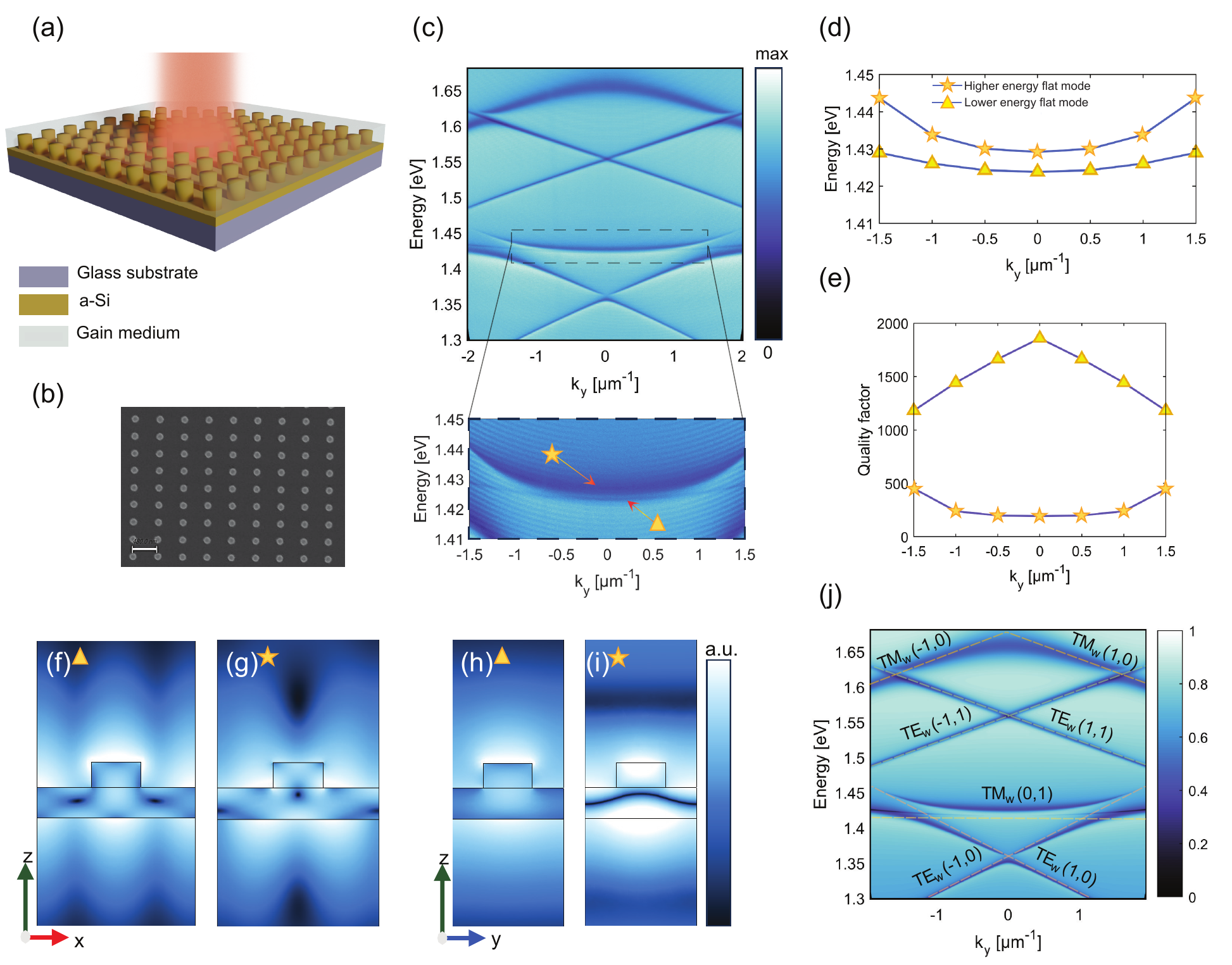}
    \caption{(a) General scheme of the samples. (b) Typical SEM image of a sample being studied. The ruler bar corresponds to 480\,nm. (c) Transmission spectrum of the sample with the periodicities of $p_x$= 480\,nm and $p_y$=\,340\,nm for unpolarized light. The inset shows two flat modes indicated by a triangle and a star in a higher spectral resolution. (d) Simulation of eigenfrequencies for the band structure of the metasurface using finite element method (FEM) in COMSOL Multiphysics for two flat modes. (e) The quality factor of two flat modes across all values of $k_y$. (f)-(i) The distributions of the absolute value of the electric field for the two flat modes within the structure's unit cell. (f) and (h) are the mode profiles of the lower energy flat mode in the zx and zy plane of view, respectively. (g) and (i) are the mode profiles of the higher energy flat mode in the zx and zy plane of view, respectively, showing high intensities also outside the waveguide layer and the nanoparticles. (j) Transmission spectrum simulated by rigorous coupled-wave analysis (RCWA) method in Lumerical RCWA solver for unpolarized light. The flat\,bands are of $TM_{w}$ type.}
    \label{fig:Fig1}
\end{figure}

In our study, we fabricated rectangular arrays of cylindrical $\alpha$-Si nanoparticles by electron beam lithography and partial etching of a thin  $\alpha$-Si waveguiding film with the following geometrical parameters: the waveguiding layer has a thickness of 85\,nm, the cylinders have a height of 70\,nm and a diameter of 150\,nm. The systems support hybrid, guided-mode resonances and feature a less dispersive, flat mode within the dispersion band. The schematic and a typical scanning electron microscope (SEM) image of the structure are shown in Figure \ref{fig:Fig1}a,b.

Figure \ref{fig:Fig1}c shows energy dispersion in the momentum space of the sample with the periodicities 480\,nm and 340\,nm in the x and y directions, respectively. The incidence plane is yz. These periodicities are chosen to show the spectral peculiarities of the system in a most illustrative way. The inset of the figure shows the flat modes in higher resolution (marked with a star and triangle). These flat modes were chosen to appear at approximately 1.43\,eV, close to the emission maximum of the dye molecules used in the lasing experiments. In addition, at the $k_y$ close to 1.7\,$\mu m^{-1}$, the observed high-energy flat mode (denoted by a star) disappears and converts to a dark state.

To have a clear insight into guided-mode resonances observed in spectra, a semi-analytical approach can be applied \cite{amedalor2023high}. The resonance occurs under phase-matching conditions between the diffraction grating of the structure and the incident wave:
\begin{equation}
\vec{k_{y}}+m_x \vec{G_x}+n_y \vec{G_y} = \vec{\beta}, \label{eqn:1}
\end{equation}
Where $\vec{k_{y}}$ is incident in-plane wave vector laying in $yz$ plane ($\vec{k_{y}}$= $k_{0}sin\Theta$),  <$m_x,n_y$> are diffraction orders, $\vec{G}_{x,y}$ are reciprocal lattice vectors, $\left|\vec{G}_{x,y}\right|$= $
\frac{2 \pi}{P_{x,y}}$, and $\vec{\beta}$ is a propagation constant, $\left|\beta\right|= n_{eff} k_{0}$, where $n_{eff}$ is an effective refractive index of a certain guided mode. To achieve a flat mode, one has to minimize the changes in $\beta$ in response to the incident angle.

\begin{figure}[hpt]
    \centering
    \includegraphics[width=16cm]{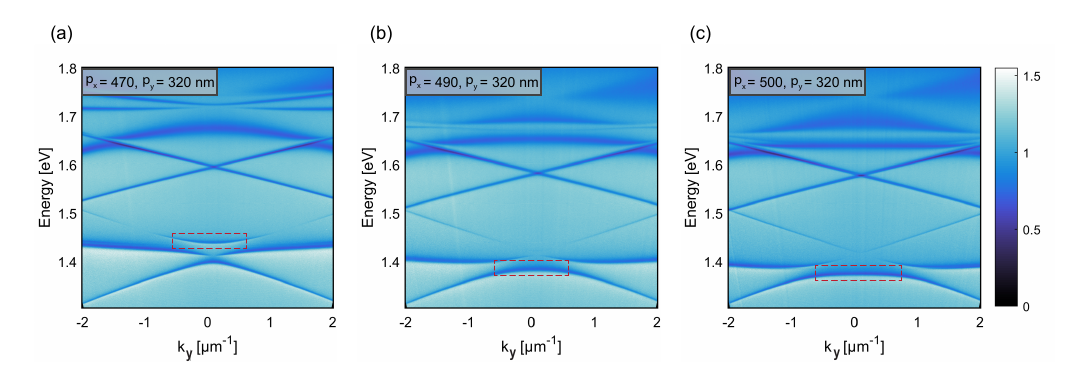}
    \caption{Experimental mode dispersion spectra affected by changing periodicity in the x direction for fixed periodicity value along the y direction, $p_y$= 320\,nm. (a) shows spectrum when $p_x$= 470\,nm. Flat mode around 1.45\,eV (indicated by the red box) is above the crossing modes and coupled so that the higher energy flat mode is extended in a shorter momentum range. (b) corresponds to an array with $p_x$= 490\,nm. In this case, flat mode (denoted by the red box) shifts to lower energies. (c) displays the spectrum corresponding to the arrays with $p_x$= 500\,nm. Here, the flat mode (marked by the red box) shifts to the lower energies and extends in a larger range of k-vectors. }
    \label{fig:fig2}
\end{figure}

The equation \ref{eqn:1} can be written in the following way:
\begin{equation}
\left(k_0 \sin \Theta+\frac{2 \pi n_{y}}{P_y}\right)^2+\left(\frac{2 \pi m_{x}}{P_x}\right)^2= \left(k_0 n_{\mathrm{eff}}\right)^2, \label{eqn:2}
\end{equation}
The flat modes dispersions are even in the in-plane momentum vector projection, therefore, $n_y= 0$. To quantify the "flatness", we introduce the ratio of the difference between mode wavelength at $k_{y}= 0$ and mode wavelength at some  $\Theta$ (for instance, maximum collection angle in the experimental setup), neglecting the dispersion of the involved refractive indices in the approximation of a flat mode:
\begin{equation}
	\frac{\lambda_{\Theta}-\lambda_{0}}{\lambda_0}=\frac{n_{sur}}{n_{eff}}|\sin(\Theta)|. \label{eqn:3}
\end{equation}
Here $n_{sur}$ is a refractive index of the surrounding medium. It can be seen that a higher effective refractive index of a guided mode leads to smaller changes in mode wavelength. 

For the lasing experiments, we consider guided modes with the <$m_x,n_y$> = <$\pm$ 1,0 > diffraction order and $TM_{w}$ polarization, where the subscript $w$ indicates that the mode is of waveguide nature.  In principle, the higher effective index of the $TE_{w}$ polarized modes makes them likely to deliver the most dispersionless modes. However, to obtain lasing, we need the electromagnetic field to be also localized in the emissive layer for efficient light amplification. This can be achieved efficiently only in $TM_w$ modes in our system (the electromagnetic field distributions for $TM_w$ and $TE_w$ modes are shown in Figure \ref{fig:Fig1} and the Supporting Information, section S1). Therefore, we focus on $TM_w$ modes.

The thickness of the waveguiding layer is carefully adjusted to control the effective refractive index of the modes. Furthermore, the extension of the flat mode to the larger k-vectors is achieved through the manipulation of periodicities in different directions and control of the waveguiding layer thickness. The periodicities of the arrays in the x and y directions were intentionally chosen to differ from each other to spectrally separate modes of the same polarization with the order of <0,m> (flat modes) and <m,0>, at least at the $\Gamma$-point (and ideally with the whole range of available values of $k_{y}$). Figure \ref{fig:fig2} shows how flat modes are affected by changing the periodicity in the x direction.

 The rigorous coupled-wave analysis (RCWA) mode dispersion simulation (Figure \ref{fig:Fig1}j) and the band structure of the two flat modes (Figure \ref{fig:Fig1}d) show good agreement with the experimental spectrum for the structure with $p_x$= 480\,nm and $p_y$= 340\,nm. However, there is a difference: the lower energy flat mode depicted in Figure \ref{fig:Fig1}j is non-radiative (discontinuous at the $\Gamma$-point region while the experimental one in Figure \ref{fig:Fig1}d is continuous). This is evident also from the quality factors of the two flat modes illustrated in Figure \ref{fig:Fig1}e. The lower energy flat mode ($f_{1}$) exhibits the highest quality factor at the $\Gamma$-point, which diminishes as the wave vector moves away from this point. This reduction can be attributed to an increase in radiation loss at larger wave vectors. The difference between the simulated and experimental spectra is likely due to symmetry-breaking introduced in the fabrication process.   
 Figure \ref{fig:Fig1}f-i displays the absolute value of the electric field spatial distributions at flat mode frequencies, \(|E\mid=\sqrt{E_x^2+E_y^2+E_z^2} \), under normal incidence. In these modes, the electric field is localized not only in a waveguiding layer but also in the emissive medium. This characteristic makes the flat modes effective for lasing experiments as they can couple more efficiently with the emitter material as compared to, for instance, $TE_w$ modes (see section S1 in the Supporting Information).

In our lasing experiment, we used a fluorescent dye IR-140 as a gain medium dissolved in a 1:2 ratio mixture of dimethyl sulfoxide (DMSO) and benzyl alcohol (BA) with a concentration of 10\,mM to operate within the weak light-matter coupling regime. The system is pumped by a vertically (along the y-direction) polarized femtosecond pulse laser (Coherent Astrella) radiation with a central wavelength of 800\,nm, a pulse duration of 100\,fs, and a pulse repetition rate of 1\,kHz. The experimental setup allowed us to obtain angle-resolved spectra of lasing radiation, in addition to real and Fourier space emission patterns. 

\begin{figure}[hpt]
	\centering
	\includegraphics[width=16cm]{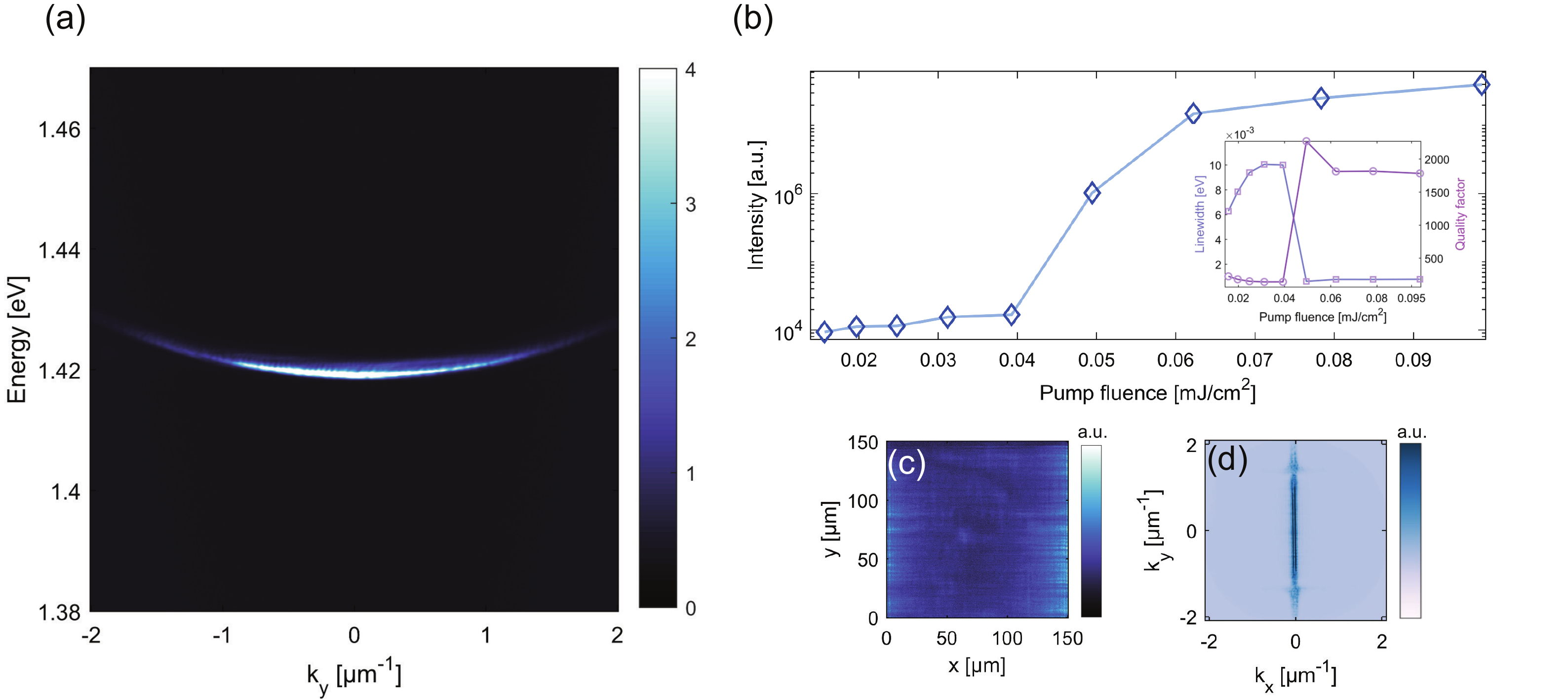}
	\caption{Lasing in the array with the $p_x$= 480\,nm and $p_y$= 340\,nm sample. The pump fluence for panel (a),(c) and (d) is 0.078\,$ \mathrm{mJ/cm^2} $.
 (a) Momentum resolved lasing emission spectrum. (b) Emission power dependence on pump fluence, inset shows the full-width-half-maximum (FWHM) of the emission (squares) and quality factor (circles) with increasing pump fluence. (c) Real and (d) momentum-space of lasing emission patterns.} 
    
	\label{fig:Fig3}
\end{figure}

\begin{figure}[hpt]
    \centering
    \includegraphics[width=16cm]{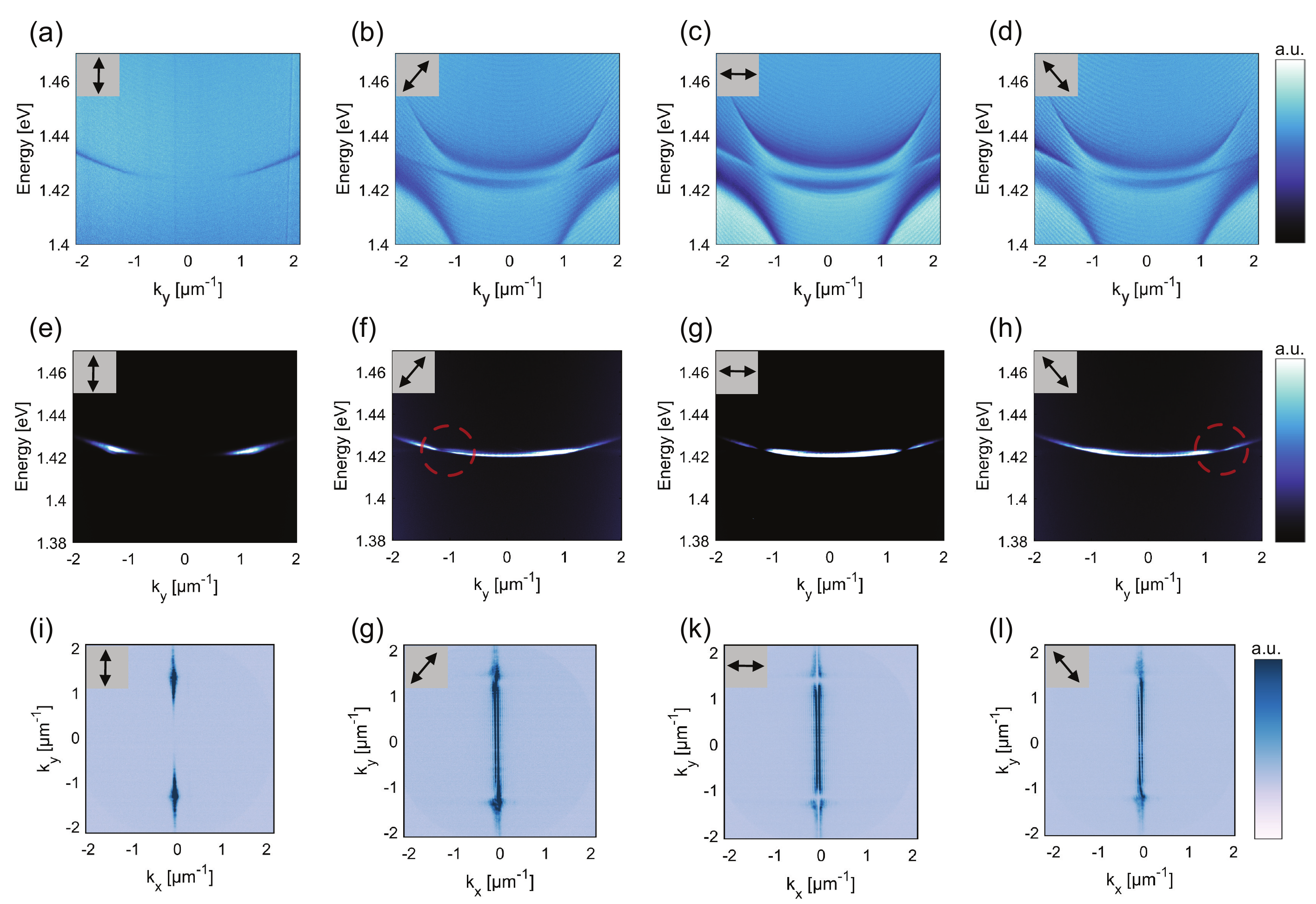}
    \caption{Polarization vorticity of accidental BIC modes in transmission and lasing spectra. (a)-(d) Energy-momentum dispersion measurements excited by a wide-range spectrum source for different polarizations denoted by arrows. (e)-(h) Polarization-resolved lasing emission spectra of the system under pump fluence of 0.098\, $ \mathrm{mJ/cm^2} $. The data was captured slightly away from the $\Gamma$-point within the region of $k_{x}<0$ (see the 2D k-space images shown in panel (i)-(l)). (i)-(l) Momentum space patterns of lasing radiation at different emission polarization states.}
    \label{fig:Fig4}
\end{figure}

The array with the periodicities of $p_x$= 480\,nm and $p_y$= 340\,nm showed lasing emission with an energy close to 1.42\,eV with a quality factor of 1816 (Figure \ref{fig:Fig3}). The threshold of the lasing is at a pump fluence of 0.04 mJ/cm$^{2}$ and relatively low compared to similar systems of plasmonic nanoparticle arrays with the same gain material~\cite{heilmann2022quasi,salerno2022loss,arjas2024high}. In a flat\,band, any spatial pattern is possible as a combination of degenerate momenta, and the lasing will occur in a configuration that optimizes the balance of gain and losses. The real-space emission distribution above the lasing threshold in Figure \ref{fig:Fig3}c displays emission extended over the whole sample, but is strongest at the edges. This looks similar to the typical pattern for dark mode lasing~\cite{hakala2017lasing}, and is connected to the dark-mode feature in the $x$ direction (no emission at exactly $k_x=0$, see also Supporting Information section S2). However, the center of the sample shows emission as well (as clearly seen in polarization analysis, Supporting Information section S3); the situation is more complex than in~\cite{hakala2017lasing} because there is emission to a broad range of $k_y$.

Figure \ref{fig:Fig4} illustrates the polarization vorticity of accidental BIC modes within both transmission and lasing spectra. In the emission path, we employ a linear polarizer set at various angles, as indicated by the double-headed arrows in the inset. The arrows pointing vertically and horizontally denote the $TM_{e}$ polarization and $TE_{e}$ polarization of light emitted to the detectors, respectively. The diagonal arrows represent diagonal polarization states oriented in different directions, both in a clockwise and counterclockwise manner.

 Figure \ref{fig:Fig4}a-d shows the mode dispersion of the array excited by a wide-range spectrum source for linear polarization rotated at different angles. It can be seen that the dispersion curve in Figure \ref{fig:Fig4}a shows signatures (dark area around the $\Gamma$-point) of symmetry-protected BIC for $TM_{e}$ polarization of light (vertical arrow in the inset). Both angle-resolved spectrum and momentum space emission pattern correspond to typical BIC lasing case, see Figure \ref{fig:Fig4}e,i~\cite{heilmann2022quasi, BIC_narrow}. 

For the diagonal and $TE_{e}$ polarization of light (diagonal and horizontal arrows), the transmission spectrum is different: the symmetry-protected BIC mode at the $\Gamma$-point disappears while two flat modes emerge; the lower one was not seen in the simulations so we attribute it to a symmetry-breaking in the fabrication process. The lower energy flat mode demonstrates two off-$\Gamma$ point BIC modes (at $k_y= 1.3\,\mu m^{-1}$), see Figure \ref{fig:Fig4}b-d, clearly visible in lasing spectra (Figure \ref{fig:Fig4}e-h). By comparing the dispersions of Figure \ref{fig:Fig4}a-d with the emission in Figure \ref{fig:Fig4}e-l we conclude that lasing happens in the lower energy modes with the symmetry-protected and accidental BICs. This is consistent with the simulations of Figure \ref{fig:Fig1}e, which show a much lower Q-factor for the higher energy mode.

The polarization vorticity usually accompanying BIC states can be seen in the momentum space patterns (Figure \ref{fig:Fig4}i-l): asymmetry with respect to $k_x= 0$ for the diagonal polarization states near areas of accidental BIC existence indicates a non-zero topological charge (|q|= 1)~\cite{zhen2014topological,doeleman2018experimental,salerno2022loss,arjas2024high,zhang2018observation, jin2019topologically}. The same polarization vorticity can also be seen in energy-momentum space lasing (Figure \ref{fig:Fig4}e-h), where only a part of the momentum space pattern corresponding to negative values of $k_x$ was directed into the spectrometer and analyzed. Energy-momentum pattern for positive $k_x$ values is antisymmetric to that one for negative $k_x$ values for diagonal polarizations. While analyzing the whole momentum-space pattern with a wider spectrometer slit, no asymmetry is found (see Supporting Information section S4).

\begin{figure}[hpt]
	\centering
	\includegraphics[width=16cm]{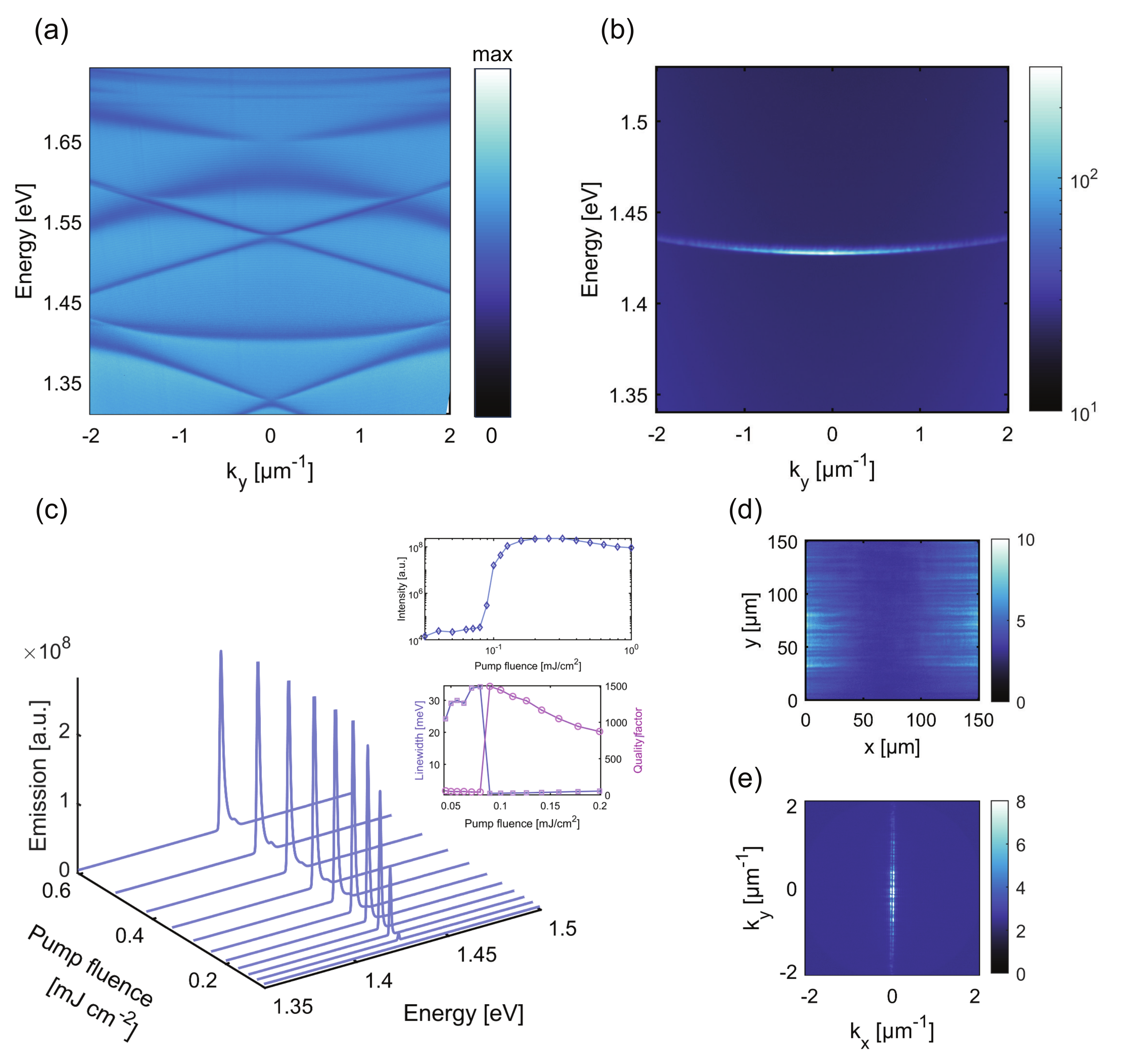}
	\caption{(a) Transmission spectrum of the sample with the periodicities $p_x$= 475\,nm and $p_y$= 350\,nm. (b) Lasing emission in a flat mode of the sample in the presence of a fluorescent dye solution (IR140, 10\,mM). (c) Sample emission spectra for different pump fluence values, the inset shows the dependence of the intensity (upper panel) of sample emission on pump fluence, averaged over $k_y$ values, and FWHM and quality factor (lower panel) of the emission representing the variation of the emission linewidth (squares) and quality factor (circles) with increasing pump fluence. Lasing emission patterns in (d) real and (e) momentum space.}
  
	\label{fig:Fig2}
\end{figure}

Figure \ref{fig:Fig2} shows the lasing emission of a sample with the lattice periodicities of 475\,nm and 350\,nm in the x and y directions, respectively; for these parameters, the flat feature was found to extend over the largest momentum range achieved in this study. The angle-resolved spectra of this structure are shown in Figure \ref{fig:Fig2}a. The flat modes appeared at 1.432\,eV, exhibiting an overlap with each other. Figure \ref{fig:Fig2} illustrates the appearance of a lasing peak with the pump fluence increase. A transition to lasing emission is evident from the abrupt change in intensity.  Lasing occurs at 1.437\,eV in a pump fluence of 0.09\,$mJ/cm^2$.

  Figure \ref{fig:Fig2}d,e depict the far-field lasing emission in real space and 2D momentum space above the lasing threshold. The general features resemble those of the  $p_x$= 480\,nm and $p_y$= 340\,nm sample described previously, but in this case, accidental BIC modes correspond to large values of $k_y$ outside the lasing region (see Supporting Information section S5), resulting in a more uniform polarization pattern in the considered $k_y$ range.

  \section*{Conclusions}

We have fabricated a rectangular metasurface with a waveguiding layer made of $\alpha$-Si, which supports tunable flat modes. Changing the geometrical parameters of the arrays (e.g., periodicity) allows fine-tuning of the flat modes, for example, their spectral position and hybridization with other modes. We achieved lasing in the flat mode: by carefully tuning the height of the guiding layer and the structure periodicities in the x and y directions, we were able to achieve near-dispersionless lasing over a large range of k-vectors up to $k_y$= $ \pm 2\,\mu m^{-1}$, with an energy variation of 0.00924\,eV. Our results demonstrate that lasing in a flat\,band of a long-range-coupled photonic system is feasible and that lasing emission to a wide angle range with a narrow energy distribution is possible. Additionally, some structures showed off-$\Gamma$ point BIC features at the flat mode. We observed a polarization vorticity corresponding to |q|= 1 charge at these momentum points in the lasing regime. As shown by the accidental BICs found, topological features can be realized as well.

Our design utilizes the intrinsically localized guided modes and combines them with a periodic structure, which allows precisely controlling the frequency and polarization of the lasing light and tailoring modes that overlap with the gain medium efficiently. These metasurfaces thus offer a highly tunable, low-loss platform for the creation of sophisticated lasing devices both in terms of polarization and direction of the emission. They offer prospects for future studies where the high density of states of flat\,bands can be used for enhancing light-matter interactions and non-linear effects, as well as lowering the lasing and polariton condensation thresholds. Furthermore, the freedom of creating spatial patterns offered by flat\,bands could be explored, for example, by the design of the gain material, defects, or non-linearities.

\section*{Materials and Methods}

\subsection*{Sample Preparation}
We fabricated an $\alpha$-Si film on a borosilicate glass substrate using the plasma-enhanced chemical vapor deposition (PECVD) method. A high resolution, negative electron beam resist (AR7520.07) layer was spin-coated and baked at \( 85^{\circ} \mathrm{C} \) for 1 minute. Next, we use electron beam lithography to pattern a $150 \times 150 \,\mu\mathrm{m^2}$
 arrays of nanocylinders with a diameter of 150\,nm on the $\alpha$-Si film. After developing the resist, the sample is partially etched through the reactive-ion etching technique (RIE), where the resist is used as a hard mask. Finally, the resist is removed by an oxygen ashing process (see section S6 in the Supporting Information). Throughout the process, atomic force microscopy (AFM) and spectroscopic ellipsometer are used to control the thickness of the waveguide film and cylinder height. In dispersion measurements, samples are immersed in an index-matching oil and covered by a microscope coverslip. The refractive index of the index-matching oil corresponds to the ones of the fluorescent dye solution and glass substrate. For the lasing measurements, the fluorescent dye solution (IR-140 dissolved in 1:2 dimethyl sulfoxide (DMSO) and benzyl alcohol (BA) mixture) with a concentration of 10\,mM is injected into a chamber created on the sample. This concentration of the fluorescent dye solution is made to provide a weak light-matter coupling regime \cite{torma2014strong}.

\subsection*{Measurement Set-up}

\begin{figure}[hpt]
    \centering
    \includegraphics[width=15cm]{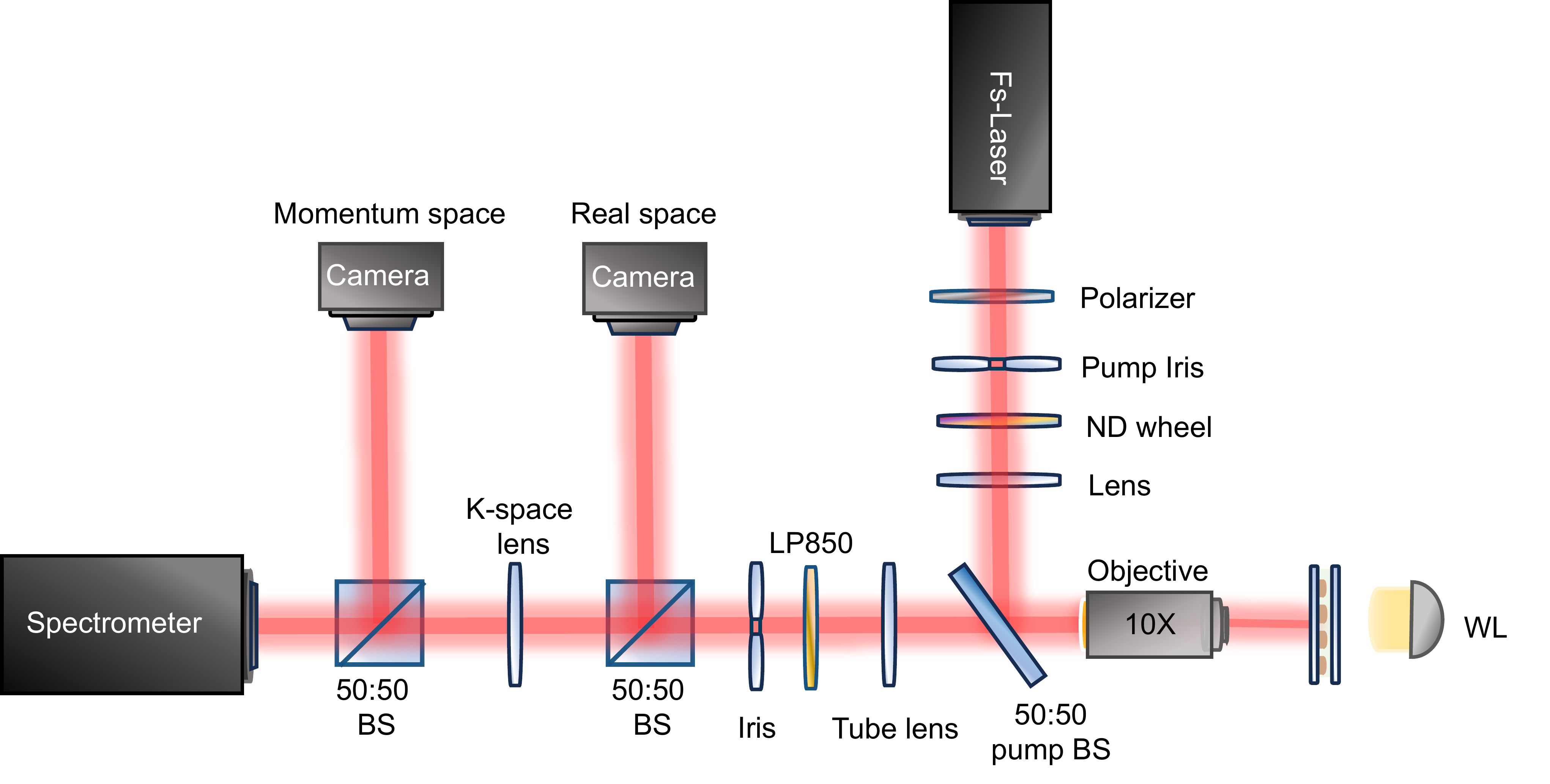}
    \caption{ Experimental set-up.}
    \label{fig:Sup2}
\end{figure}

We used an angle-resolved spectroscopy setup for our transmission and lasing measurements. The set-up schematic is depicted in Figure \ref{fig:Sup2}.  A dry objective (NA= 0.3, 10X) was utilized to gather the light from the sample, providing a maximum light collection angle of approximately 17 degrees. After the objective, the light passed through a tube lens with a focal distance of 200\,mm. For lasing experiments, a long-pass filter with a cut-off wavelength of 850\,nm (1.46\,eV) was used to suppress pump radiation. Two CMOS cameras were used to capture sample emission images in real and momentum space. Then, the light was analyzed using a spectrometer resolving both the light spectrum and its distribution for different $k_y$ values in the range corresponding to the numerical aperture of the objective ($k_y^{max}=\frac{2\pi}{\lambda}NA$, where $\lambda$ is a free space wavelength). A femtosecond Ti:Sapphire laser (Coherent Astrella) with the following parameters was used as a pump laser: pulse repetition rate 1\,kHz, pulse width 100\,fs, central wavelength 800\,nm (1.55\,eV). An additional bandpass filter with a central wavelength of 800\,nm and a bandwidth of 40\,nm was set in the pump channel to effectively separate pump laser radiation and lasing emission from the sample. Laser radiation was spatially cropped by an iris, its image was transferred to the sample plane by an optical system consisting of a lens and the objective providing uniform pump intensity on the arrays. The pump laser was linearly polarized along the y-direction. For transmission measurements, a halogen lamp was used as a light source. For polarization-resolved measurements, a thin film polarizer was placed in the signal channel.

\section*{Funding Sources}

This work was supported by the Research Council of Finland under Project Nos.~349313 and 318937 (PROFI) and by the Jane and Aatos Erkko Foundation and the Technology Industries of Finland Centennial Foundation as part of the Future Makers funding program. Funded by the European Union. Views and opinions expressed are, however, those of the author(s) only and do not necessarily reflect those of the European Union or the European Innovation Council and SMEs Executive Agency (EISMEA). Neither the European Union nor the granting authority can be held responsible for them. (SCOLED, Grant Agreement No. 101098813).
The work is part of the Research Council of Finland Flagship Programme, Photonics Research and Innovation (PREIN), decision number 346529, Aalto University. Part of the research was performed at the OtaNano Nanofab cleanroom (Micronova Nanofabrication Centre), supported by Aalto University. We acknowledge the computational resources provided by the Aalto Science-IT project.

\begin{acknowledgement}
We thank Kristian Arjas for the helpful scientific discussions. 
\end{acknowledgement}

\section*{Supporting Information}
The following files are available free of charge.
\begin{itemize}
  \item Filename: Electric field distribution for $TE_w$ flat mode (S1); Correlation between flat\,band and dark modes in different planes of incidence (S2); Polarization vorticity feature in real-space images (S3); Characteristics of polarization vorticity in lasing emission within energy-momentum spectra (S4); Polarization vorticity feature in an array hosting extended flat mode in momentum-space (S5); Sample fabrication process (S6).
\end{itemize}

\bibliography{MyRef}

\end{document}